\documentclass[%
 reprint,
 amsmath,amssymb,
aps,
pra,
superscriptaddress,
longbibliography,
]{revtex4-2}

\usepackage{graphicx}
\usepackage{dcolumn}
\usepackage{bm}
\usepackage{hyperref}
\usepackage{color}
\usepackage{tabularx}
\usepackage{float}
\usepackage{upgreek}
\usepackage{nicefrac} 
\usepackage{enumitem}
\usepackage{placeins}
\usepackage{tabularx}
\usepackage{braket}
\usepackage{acronym}
\usepackage{tikz}
\usepackage{comment}
\usepackage[capitalise]{cleveref}

\specialcomment{notes}{\begingroup\sffamily\color{blue}}{\endgroup}

\newsavebox{\foobox}

\newcommand{\TwoEightSi}{\ensuremath{^{28}\text{Si}}}
\begin{document}

\title{Memory and transduction prospects for silicon $T$ centre devices}

\author{Daniel B. Higginbottom}
\affiliation{Department of Physics, Simon Fraser University, Burnaby, British Columbia V5A 1S6, Canada }
\affiliation{Photonic  Inc., Coquitlam,  BC,  Canada}
\author{Faezeh Kimiaee Asadi}
\affiliation{Institute for Quantum Science and Technology, and Department of Physics \& Astronomy, University of Calgary, 2500 University Drive NW, Calgary, Alberta T2N 1N4, Canada}
\author{Camille Chartrand}
\affiliation{Department of Physics, Simon Fraser University, Burnaby, British Columbia V5A 1S6, Canada }
\affiliation{Photonic  Inc., Coquitlam,  BC,  Canada}
\author{Jia-Wei Ji}
\affiliation{Institute for Quantum Science and Technology, and Department of Physics \& Astronomy, University of Calgary, 2500 University Drive NW, Calgary, Alberta T2N 1N4, Canada}
\author{Laurent Bergeron}
\affiliation{Department of Physics, Simon Fraser University, Burnaby, British Columbia V5A 1S6, Canada }
\affiliation{Photonic  Inc., Coquitlam,  BC,  Canada}
\author{Michael L. W. Thewalt}
\affiliation{Department of Physics, Simon Fraser University, Burnaby, British Columbia V5A 1S6, Canada }
\author{Christoph Simon}
\affiliation{Institute for Quantum Science and Technology, and Department of Physics \& Astronomy, University of Calgary, 2500 University Drive NW, Calgary, Alberta T2N 1N4, Canada}
\author{Stephanie Simmons}
\affiliation{Department of Physics, Simon Fraser University, Burnaby, British Columbia V5A 1S6, Canada }
\affiliation{Photonic  Inc., Coquitlam,  BC,  Canada}

\setlength{\skip\footins}{0.5cm}

\begin{abstract} 
The $T$ centre, a silicon-native spin-photon interface with telecommunications-band optical transitions and long-lived microwave qubits, offers an appealing new platform for both optical quantum memory and microwave to optical telecommunications band transduction. A wide range of quantum memory and transduction schemes could be implemented on such a platform, with advantages and disadvantages that depend sensitively on the ensemble properties. In this work we characterize $T$ centre spin ensembles to inform device design.
We perform the first $T$ ensemble optical depth measurement and calculate the improvement in centre density or resonant optical enhancement required for efficient optical quantum memory. We further demonstrate a coherent microwave interface by coherent population trapping (CPT) and Autler-Townes splitting (ATS). We then determine the most promising microwave and optical quantum memory protocol for such ensembles. By estimating the memory efficiency both in free-space and in the presence of a cavity, we show that efficient optical memory is possible with forecast densities. Finally, we formulate a transduction proposal and discuss the achievable efficiency and fidelity. \end{abstract}

\maketitle

\section{Introduction}
Quantum memories and transducers are enabling technologies for the global quantum internet. Optical quantum memories that coherently store and recall unknown states of a traveling optical field on demand \cite{Lvovsky2009} are a component of quantum repeaters \cite{briegel1998quantum,childress2005fault, sangouard2011quantum, asadi2018quantum} to enhance the entanglement distribution range, and therefore the secure communication range, of quantum networks \cite{kimble2008quantum} as well as a means of synchronizing processes with optical quantum computers. Microwave quantum memories, on the other hand, can interface with microwave frequency material qubits, including superconducting circuit qubits---arguably the most advanced quantum computing platform at present---to extend circuit depth and reduce no-op error rates \cite{PhysRevLett.127.140503}. 

Long-coherence spin ensembles have been proposed as quantum memories for both applications, with considerable research into optical quantum memories in particular. To date, several quantum memory protocols including electromagnetically induced transparency (EIT) \cite{fleischhauer2005electromagnetically}, atomic frequency comb (AFC) \cite{afzelius2009multimode}, controlled reversible inhomogeneous broadening (CRIB) \cite{kraus2006quantum}, and Autler-Townes Splitting (ATS) \cite{Saglamyurek2018a} have been proposed based on platforms such as rare-earth ions \cite{ma2021one, hedges2010efficient}, NV centres in diamond \cite{heshami2014raman}, atomic vapours \cite{guo2019high, katz2018light}, and Bose-Einstein condensates (BEC) \cite{zhang2009creation, saglamyurek2021storing}. However, few of these systems operate in a telecommunications (telecom) band. Extending the quantum internet to global scale requires optical quantum memories compatible with satellite communications as well as terrestrial optical fibre networks.

Similarly, efficient and coherent conversion between gigahertz microwave and terahertz optical signals is essential for networking microwave material qubits \cite{Lauk2020}. Transduction from microwave to telecom photons could enable, for example, networking superconducting quantum processors and spin quantum processors \cite{Vandersypen2019,Xue2022} via the global quantum internet. There are several remarkable experimental demonstrations \cite{HaiTao2022, higginbotham2018harnessing, fan2018superconducting, vogt2019efficient} as well as theoretical approaches \cite{obrien2014, li2017quantum, asadi2022proposal} on reversibly converting between microwave and optical fields. In general, quantum transduction can be carried out through intermediate systems such as atomic ensembles \cite{obrien2014,Fernandez-Gonzalvo2015, HaiTao2022,vogt2019efficient}, electro-optical systems \cite{soltani2017efficient,fan2018superconducting}, electro-optomechanical \cite{stannigel2010optomechanical, higginbotham2018harnessing}, and others \cite{hisatomi2016bidirectional, das2017interfacing}. 

Ensembles of atoms or atom-like systems combining long-coherence microwave spin qubits and optical transitions are therefore a versatile `swiss-army knife' platform suitable for microwave and optical memory as well as microwave to optical transduction. Identifying and developing long-coherence quantum systems combining narrow and absorptive microwave and telecom optical transitions is an important milestone for designing a quantum network. However, to date few suitable systems have been identified. Memory and transduction proposals of this kind have focused almost exclusively on Er$^{3+}$, which combines narrow MW and telecom optical transitions in many crystal hosts \cite{Williamson2014a,obrien2014,Fernandez-Gonzalvo2015}. Although the optical transition is comparatively weak, significant optical depths have been observed with concentrated Er$^{3+}$ ensembles in large crystals \cite{xie2021}. 

The newly rediscovered silicon $T$ centre offers native O-band telecom access and long-lived spins operating at microwave frequencies \cite{Bergeron2020, higginbottom2022}. Notably, silicon is an attractive host material for such a system, as it is already the substrate of choice for a wide variety of quantum platforms including ion traps~\cite{Cho2015}, quantum dots~\cite{Veldhorst2017}, superconducting qubits~\cite{Arute2019}, and, naturally, impurity and defect spins in silicon~\cite{Pla2012} and silicon photonics~\cite{Wang2020}. The $T$ centre integrates with silicon photonic devices on CMOS-compatible wafers as a waveguide and fibre-networked quantum information platform \cite{higginbottom2022} that could further mediate between other silicon quantum technologies packaged on-chip, such as superconducting, trapped ion and gate-defined quantum dot qubits, for a complete, all-silicon hybrid quantum information platform.

In this paper we introduce the prospects of $T$ centres for microwave quantum memory, optical quantum memory, and microwave-optical quantum transducers. We will first briefly review the known properties of the $T$ centre as revealed by recent studies. Then, we further characterize $T$ centre ensembles experimentally towards the goal of quantum memories and transducers. We measure $T$ centre optical depth (OD) in \cref{sec:OD}, microwave coherent population trapping (CPT) in \cref{sec:CPT}, and microwave Autler-Townes splitting in \cref{sec:ATS}. Informed by these measurements, in \cref{sec:memory}, we estimate the efficiency of applicable optical memory schemes including EIT and ATS. In \cref{sec:mw} we derive the resonator requirements for efficient $T$ centre microwave quantum memories. In \cref{sec:transduction}, we consider the potential of combining these ATS schemes to design a transduction protocol. Finally, we conclude with future directions in \cref{sec:discussion}.

\section{The $T$ centre}
\label{sec:t_review}
The $T$ centre is a luminescent silicon defect with a sharp emission line at $935$~meV, in the telecommunications O band. Early work \cite{Irion1985a,Safonov1996b} established that $T$ centres comprise two nonequivalent carbon atoms and a hydrogen atom. Some of the $T$ centre parameters required to calculate quantum memory and transduction efficiencies, including the level structure, optical linewidths and lifetimes, and spin coherence times have recently been established \cite{Bergeron2020,MacQuarrie2021,higginbottom2022}. The remaining key parameters---optical depth, microwave coupling strengths and two-photon microwave linewidth---are obtained in this work and put to use in proof-of-principle quantum memory precursor experiments. 

Spectroscopy of $T$ centre ensembles \cite{Bergeron2020} established that the ground state has a tightly bound s\,=\,1/2 electron coupled to the hydrogen nuclear s\,=\,1/2 spin through an anisotropic hyperfine interaction. In its optically excited state a bound exciton is formed, the electron spins form a singlet state, and there remains an unpaired s\,=\,3/2 hole. The four-fold degeneracy of free holes in silicon is lifted in the reduced symmetry of the $T$ centre to form two distinct excited state doublets labeled TX$_0$ and TX$_1$ separated by $1.76$~meV. Thermal excitation between these two states is negligible below $\sim2$~K. Studies of TX$_0$ with ensembles and single centres revealed remarkable optical linewidths sufficient to resolve the ground state electron spin splitting at low fields \cite{Bergeron2020,higginbottom2022}. 

The TX$_0$ lifetime is 0.94~$\upmu$s in bulk silicon \cite{Bergeron2020}. The radiative efficiency, $\eta_\mathrm{R}$, of this transition is not precisely known. To date, measurements with bulk ensembles have not found evidence of non-radiative relaxation \cite{Bergeron2020}. First-principles theoretical calculations indicate $0.19 < \eta_\mathrm{R} < 0.72$ \cite{Dhaliah2022} and single-centre photon fluorescence rates bound $\eta_\mathrm{R} \geq 0.03$ \cite{higginbottom2022}.  The zero-phonon fraction, or Debye-Waller (DW) factor, is known to be $\eta_\mathrm{DW}=0.23$ \cite{Bergeron2020}.

\begin{figure}[]
\includegraphics[width=\linewidth]{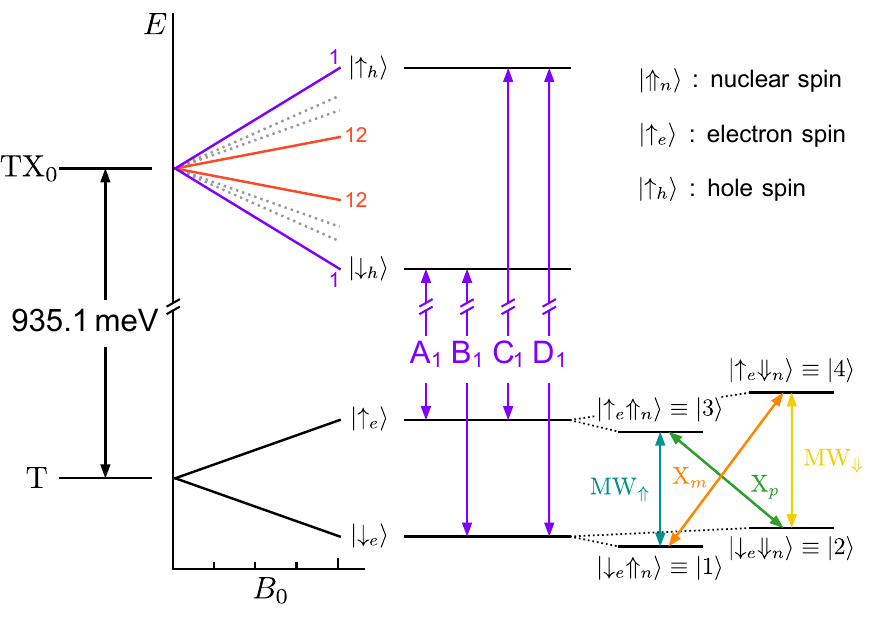}
\caption{Schematic energy level diagram of the $T$ centre 935.1\,meV TX$_0$ optical transition. Under magnetic field, the unpaired electron spin of the paramagnetic ground state $T$ and unpaired hole spin of the excitonic excited state TX$_0$ split into doublets. The anisotropic hole spin determines 12 orientational subsets for a given magnetic field orientation, with optical transitions A$_i$--D$_i$ for subset $i$ that may be optically resolved. The nuclear spin transitions are resolvable with microwave addressing.}
\label{fig:1}
\end{figure}

The schematic level structure in \cref{fig:1} shows TX$_0$ under an applied magnetic field, with four resolved optical transitions A--D between the ground state electron and excited state hole states. The $T$ centre has 24 possible orientations forming 12 inequivalent inversion symmetric subsets $i$, each with its own set of optical transitions A$_i$--D$_i$ determined by the effective anisotropic hole Land\'{e} factor $0.85 < g_{\mathrm{H}i} < 3.5$ for a particular magnetic field orientation. Optically detected magnetic resonances (ODMR) reveal ground state hyperfine splitting under microwave or RF excitation. Both the `allowed' nuclear spin preserving microwave transitions, MW$_\mathrm{\Uparrow}$ and MW$_\mathrm{\Downarrow}$, and the `forbidden' microwave transitions, X$_\mathrm{m}$ and X$_\mathrm{p}$, are observed to be strong. The $T$ centre therefore boasts both optical and microwave $\Lambda$ and V transitions for interfacing between travelling fields and spins. The nuclear spin states of the hydrogen nucleus, which have not yet been optically resolved, would become optically accessible in lifetime-limited homogeneous ensembles.

Isotopically purified \TwoEightSi{} improves upon both the spin and optical properties of the $T$ centre. Removing $^{29}$Si atoms (spin 1/2 nuclei) reduces the magnetic noise bath and improves native spin coherence times. Removing mass variations due to both $^{29}$Si and $^{30}$Si dramatically reduces the optical inhomogeneous broadening. Ensemble optical inhomogeneous linewidths as low as $33$~MHz have been observed in \TwoEightSi{} $T$ centre ensembles \cite{Bergeron2020}. Measurements of spectral diffusion have indicated that the \TwoEightSi{} ensemble optical linewidths may be close to diffusion limited, and that spectral diffusion remains low in natural silicon samples even as the inhomogeneous linewidth increases to $9$~GHz \cite{MacQuarrie2021}. Hahn echo measurements in \TwoEightSi{} crystals revealed $T$ centre electron and nuclear spin coherence times ($T_2$) of $2.1$~ms and $1.1$~s respectively \cite{Bergeron2020}. Both $T$ centre spins offer a memory advantage compared to superconducting qubits, and the nuclear spin coherence time is furthermore sufficient for repeaters in terrestrial quantum networks.

In the following sections we extend these studies by characterizing the optical absorption properties of \TwoEightSi{} $T$ centre ensembles to determine the requirements for $T$ centre optical quantum memories. We further implement the first coherent $\Lambda$ and V schemes for $T$ centres as a precursor to quantum memory and transduction protocols.

\section{Optical depth}
\label{sec:OD}

\begin{figure}[]
\includegraphics[width=\linewidth]{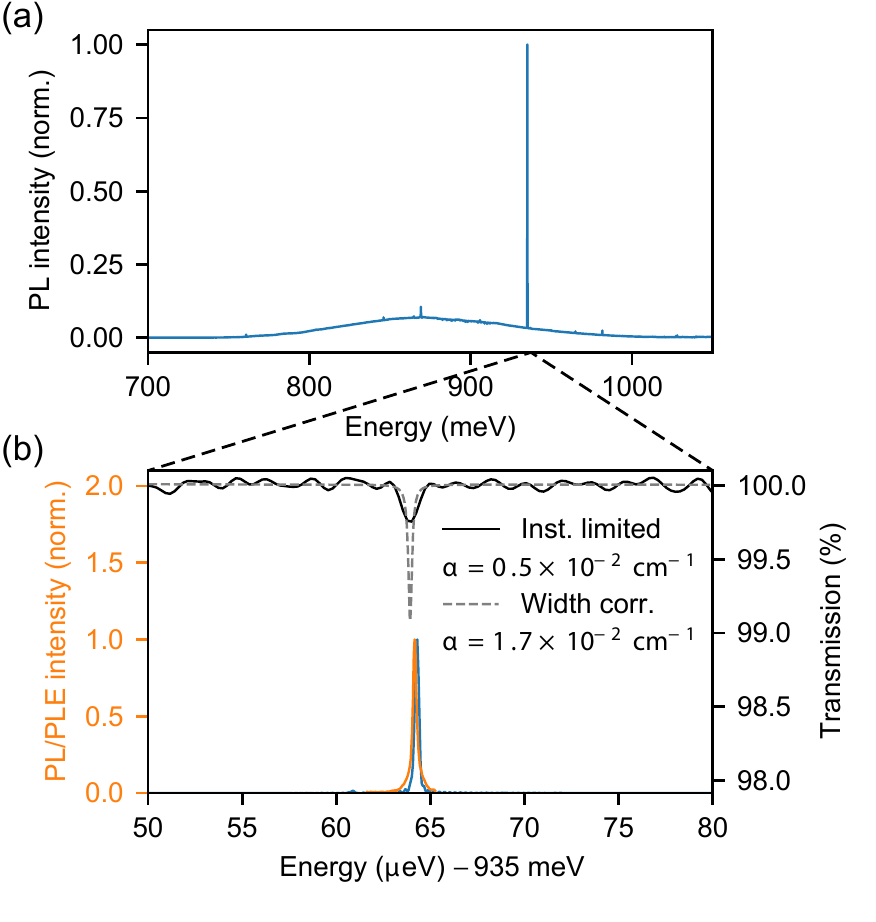}
\caption{(a) PL spectrum of a $T$ centre ensemble in \TwoEightSi. (b) High-resolution PL spectrum about the TX$_0$ zero phonon peak (blue) and PLE spectrum of the same sample (orange). The corresponding absorption spectrum (black) is instrument resolution limited. We calculate an area-preserving corrected absorption spectrum (dashed black) with linewidth and shape matched to the PLE spectrum.}
\label{fig:5}
\end{figure}

An efficient quantum memory fundamentally requires signal absorption approaching 100\% for the reversible transfer of information. Here we report the first optical absorption measurements of $T$ centres and extrapolate what optical depths are feasible within the near term for $T$ centre quantum memories. An $\approx5$~mm thick \TwoEightSi{} sample was electron-irradiated and annealed according to the procedure in \cref{sec:Methods} to produce a significant concentration of $T$ centres. We measure the complete photoluminescence (PL) spectrum under above-band excitation with Fourier transform infrared (FTIR) spectroscopy, \cref{fig:5}(a). Detail about the prominent TX$_0$ ZPL is shown in \Cref{fig:5}(b). In addition to the PL spectrum (blue), we resolve the ZPL by photoluminescence excitation (PLE) spectroscopy, resonantly exciting the TX$_0$ with a scanning laser and detecting fluorescence from the $T$ centre phonon sideband to determine an ensemble ZPL linewidth of $56$~MHz (orange). This linewidth is an inhomogeneous factor of 331 over the lifetime-limited linewidth, and slightly larger than the best measured \cite{Bergeron2020}. The corresponding FTIR optical transmission spectrum (black), taken with a continuum light source and $1.4$~K sample temperature, is presented on a separate axis of \Cref{fig:5}(b). Absorption at the $T$ centre ZPL is evident with peak absorption of 0.27(1)\%, limited by the instrument resolution. Assuming the true absorption linewidth is identical to the PL linewidth, we infer a corrected peak absorption of 0.93(3)\% corresponding to peak resonant OD $d_\mathrm{corr.} = 0.009$ and resonant absorption coefficient $\alpha_\mathrm{corr.} =  0.017$/cm.

This first absorption measurement allows us to assess the feasibility of $T$ centre ensembles in the complete absorption (OD$>5$) regime required for optical quantum memories. First, we can compare the concentration of this sample to theoretical limits and concentrations observed in other work. The $T$ centre concentration $[c]$ can be determined from the integrated absorption line area $\int \! \alpha \,\mathrm{dV}$ shown in Fig.~\ref{fig:5} \cite{Hilborn1982}

\begin{equation}
[c] = \frac{g_1}{g_2} \frac{8 \pi n^2 \tau_\mathrm{ZP} \int \alpha \mathrm{dV} }{h \lambda^2},
\label{eq:lifetime}
\end{equation}

\noindent where $g_{1,2}$ are the degeneracy of the ground and excited states, respectively, $n = 3.45$ is the silicon index of refraction at cryogenic temperatures, $\tau_\mathrm{ZP} = \tau/\left(\eta_\mathrm{R} \eta_\mathrm{DW} \right)$ is the ZP radiative lifetime, $h$ is Planck's constant, and $\lambda = 1326$~nm is the wavelength.  

Assuming the demonstrated upper (lower) bound radiative efficiency $\eta_\mathrm{R} = 1 (0.03)$ we determine the $T$ centre bulk concentration $[c] = 8.2 (2)\times10^{10}$~cm$^{-3}$ ($2.7 (0.7) \times 10^{12}$~cm$^{-3}$). This is, to our knowledge, the highest concentration of $T$ centres measured in \TwoEightSi{}. In this case concentration is limited by the very low residual carbon contamination of our sample (given in \cref{sec:Sample}). 

Concentrations up to $100$ times larger than this lower concentration bound, at least $1.7\times10^{13}$~cm$^{-3}$, have been produced in silicon photonics using a carbon implantation recipe \cite{higginbottom2022}. Orders of magnitude improvement in concentration have also been achieved using ion implantation for other radiation damage centres \cite{Murata2011,Rotem2007}. This bodes well for the feasibility of \TwoEightSi{} ensembles with considerable optical depth. We expect an order of magnitude improvement over the measured $T$ concentrations in Ref.~\cite{higginbottom2022} is possible. In the unit radiative efficiency case, a $15$~mm $^{28}$Si crystal with such a $T$ concentration will have optical depth $d=27$ suitable for efficient quantum memory. We explore the optimum memory schemes for such an ensemble in \cref{sec:memory}. It remains to be seen what additional inhomogeneous broadening is introduced in [c]$\sim 10^{14}$~cm$^{-3}$ $T$ ensembles, which could reduce the peak resonant OD. However it is also possible that further material or implant optimization will reduce the inhomogeneous linewidth, with corresponding OD improvements.

In addition to concentration or sample length improvements, optical resonators can be used to increase the resonant OD of spin-photon ensembles. Many of the most efficient transduction demonstrations leverage optical resonators \cite{higginbotham2018harnessing,fan2018superconducting}, although microwave to optical upconversion has been demonstrated with $82$\% efficiency with a cavity-free atomic ensemble \cite{HaiTao2022}. Silicon is a superb platform for monolithic resonators due to the extremely low intrinsic loss coefficient of silicon in the telecommunications bands ($1.6\times 10^{-5}$ at $1326$~nm \cite{Green2008}). Silicon Fabry-Perot and whispering gallery mode \cite{Markosyan2018} resonators have been demonstrated with finesse many orders of magnitude beyond the $F\sim1000$ required for efficient optical quantum memory with even the limited concentration of our \TwoEightSi{} $T$ sample. These resonators were manufactured from intrinsic FZ silicon, but no performance reduction is expected using \TwoEightSi{}. Even with the broader inhomogeneous linewidths that have been measured so far, high density $T$ centres implanted in natural silicon samples ($9$~GHz linewidth \cite{MacQuarrie2021}) can be enhanced to optical depth sufficient for quantum memory with only the same $F\sim1000$ low finesse cavities. 

Finally, ensembles can be incorporated into integrated silicon on insulator (SOI) silicon photonic crystal \cite{Asano2017} or waveguide ring cavities. Typical intrinsic waveguide losses in such devices are $\sim2.5$~dB/cm, imposing an upper limit on the effective ensemble length. Integrated ensembles therefore require higher concentrations than the factor of ten improvement forecast above, or narrower integrated inhomogeneous linewidths than those demonstrated in devices to date ($18$--$50$ GHz) \cite{MacQuarrie2021,higginbottom2022}. In any case, it is necessary to consider how the inequivalent $T$ centre orientational subsets will impact memory performance. Memory schemes utilizing only a single orientation will suffer an optical depth penalty as high as a factor of $12$, but possibly less depending on orientation degeneracy for a chosen magnetic field direction. This orientation penalty would be reduced by the (untested) ability to form a given orientation preferentially.

\section{Microwave coherent population trapping}
\label{sec:CPT}

As a precursor to quantum memory protocols, we next demonstrate microwave (MW) coherent population trapping (CPT) \cite{Arimondo1976} with another \TwoEightSi{} $T$ centre ensemble. CPT is a quantum phenomenon in which an equilibrium atomic superposition is prepared by two coherent electromagnetic fields \cite{Gray1978}. CPT underpins three-level coherent atom-light phenomena, including EIT \cite{Boller1991a} and stimulated Raman adiabatic passage (StiRAP). Observing an EIT window requires sufficient ensemble absorption, but the accompanying CPT can be observed even in low optical depth ensembles by ODMR spectroscopy. 
The use of optical fields to generate coherent dark states has been extensively studied \cite{Alzetta1976}. In comparison, relatively little effort has been focused on MW CPT \cite{Rakhmatullin1998,Wei1999,Childress2010,Novikov2015,Jamonneau2016}.

The optical ground state of the $T$ centre under magnetic field $B_0$ boasts two possible microwave $\Lambda$ schemes for coherently coupling the nuclear spin states $\ket{1}$ and $\ket{2}$ via the higher energy electron spin states $\ket{3}$ and $\ket{4}$ as shown in \cref{fig:1}. The $T$ centre is therefore a potential platform for $\Lambda$ microwave quantum memory schemes.

To demonstrate this capability a \TwoEightSi{} $T$ sample, prepared by the same method, is chosen for its narrow optical inhomogeneous linewidth of $33$~MHz. A static magnetic field $B_0 = 80$~mT is applied along the [110] axis of the sample, splitting the TX$_0$ transition. Under this magnetic field $10$ of the $12$ inequivalent orientational subsets $i$ produce their own set of four optically resolved transitions A$_i$, B$_i$--D$_i$ determined by the Land\'{e} $g$ tensor of the anisotropic hole. Following Ref.~\cite{Bergeron2020}, which used the same magnetic field configuration, we label the two remaining optically unresolved orientations $i=1$,$1'$. We may select distinct subsets for ODMR measurement by resonant optical excitation. For the following ODMR measurements we choose the $i=1$,$1'$ subsets. Although these subsets are not optically resolved \cite{Bergeron2020}, we will resolve the distinct orientations by CPT.

\begin{figure}[t!]
\includegraphics[width=\linewidth]{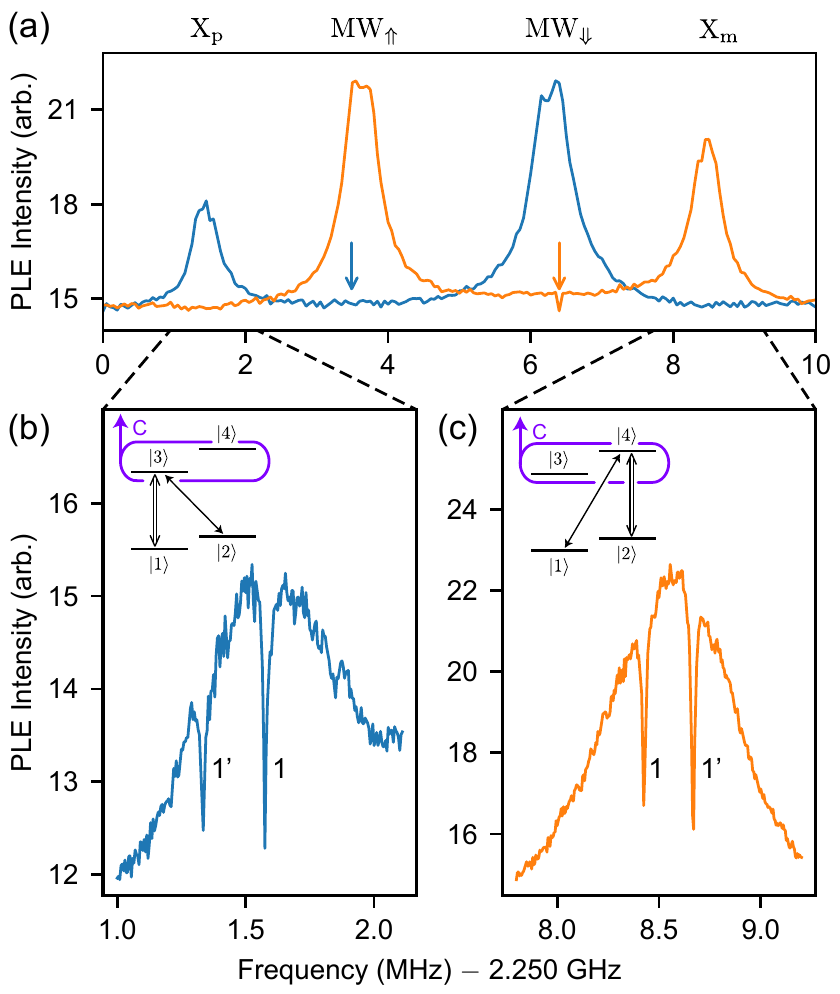}
\caption{ODMR spectra show the sideband photoluminescence as a function of MW probe detuning. Resonant optical excitation of TX$_0$ C provides electron-spin selective readout from $\ket{\uparrow_\mathrm{e}}$. (a) Two of the four resonances MW$_{\Uparrow /\Downarrow}$, X$_\mathrm{p,m}$ are visible depending on the configuration of a MW pump (pump frequency arrow matches corresponding spectra colour). The outer resonances are MW $\Lambda$ configurations. (b, c) High resolution ODMR spectra of the $\Lambda$ resonances show narrow CPT dips at two-photon resonance for each of the 1 and 1' orientation sub-ensembles. Inset Level schemes show the corresponding optical and MW configuration.}
\label{fig:2}
\end{figure} 

A `readout' laser resonant with C$_1$ selects these two optically degenerate subsets for ODMR spectroscopy. In the absence of any additional field the readout simply pumps these two optically degenerate orientational subsets into the electron spin state $\ket{\downarrow_e}$ and this hyperpolarized ensemble does not fluoresce. Adding a single pump MW field that selects only one of the two hyperfine split states is still insufficient to prevent hyperpolarization, instead pumping the system to whichever of hyperfine states $\ket{1}$,$\ket{2}$ remains unaddressed. In these ODMR measurements a MW `pump' is set resonantly with one of the four available MW transitions and a second MW `probe' scans over all four transitions. Continuous photoluminescence is detected when hyperfine states $\ket{1}$ and $\ket{2}$ are separately addressed by the MW fields. The ODMR spectra in  \cref{fig:2}(a) show two resonances per pump frequency. As expected resonances involving the `forbidden' X$_\mathrm{p,m}$ are possible, but weaker and narrower than resonances with only MW$_{\Downarrow, \Uparrow}$ (linewidths of 462(13), 506(9), 698(8) and 583(7)~kHz respectively). 

CPT is evident in each of the two possible MW $\Lambda$ configurations, shown inset to \cref{fig:2}(a). When the MW fields are in two-photon resonance CPT produces a dark state superposition of $\ket{1}$ and $\ket{2}$ that does not couple to the readout laser. Higher resolution ODMR spectra in the region of the two-photon resonances, \cref{fig:2}(b) and (c), show CPT luminescence dips with $16$($2$)~kHz linewidth---significantly narrower than the ODMR lines. For each configuration in \cref{fig:2} there are two dark resonances, one for each of the distinct orientational subsets 1 and 1' and separated by $242(1)$~kHz, below the inhomogeneous linewidth and unresolvable by above-band PL. The low-power CPT linewidths are nuclear spin coherence limited, which is already known to be longer than $1$~s at this temperature \cite{Bergeron2020}. However, at the microwave powers required to produce sufficient luminescence signal these CPT linewidths are broadened by the microwave Rabi frequencies $\Omega_{\Uparrow/\Downarrow}$, $\Omega_\mathrm{p/m}$. 

The observed ODMR and CPT profiles in \cref{fig:2} are the sum of fluorescence from the two near-degenerate orientational sub ensembles 1 and 1', in proportion determined by relative concentration and dipole orientation. The precise ODMR peak associated with each ensemble is unknown. When one ensemble is in two-photon dark resonance, fluorescence from the remaining out-of-resonance ensemble is still present, with the effect that neither CPT dip apparently achieves complete fluorescence extinction as expected in the ideal low-power case.

\begin{figure}[]
\includegraphics[width=\linewidth]{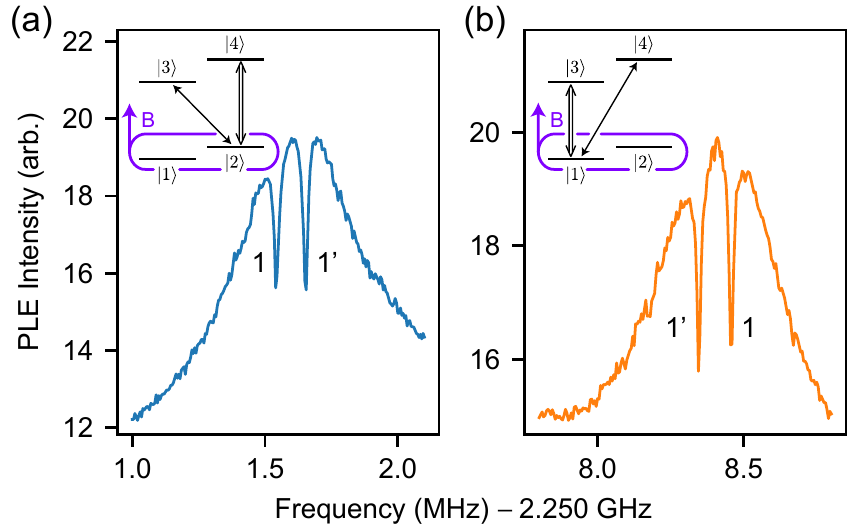}
\caption{ODMR spectra with optical readout via TX$_0$ B$_1$. In this configuration CPT is possible in each of the two V schemes, inset (a) and (b) alongside the corresponding data. Coherences are prepared in the opposite electron spin state from \cref{fig:2}}
\label{fig:3}
\end{figure}

Optically pumping transition C empties states $\ket{3}$ and $\ket{4}$, rapidly destroying any coherence $\hat{\rho}_{34} = \braket{3|\hat{\rho}|4}$, where $\hat{\rho}$ is the density matrix of the centre, and power broadening the MW transitions. However the choice to read using optical transition C is arbitrary, and we can invert the system to achieve two-photon coherence via the less common V scheme. Choosing transition B instead empties $\ket{1}$ and $\ket{2}$ and removes coherence $\hat{\rho}_{12}$. This configuration sustains nuclear spin coherences of the $\ket{\uparrow_\mathrm{e}}$ hyperfine manifold, and we therefore expect CPT in V two-photon resonance, as shown in \cref{fig:3}. The V CPT linewidths are $17(2)$~kHz, identical to the $\Lambda$ case. The pump-probe MW frequency difference at each $\Lambda$ or V CPT condition determines effective hyperfine constants $-2.93(1)$~MHz and $-2.57(1)$~MHz for the 1 and 1' sub-ensembles, respectively.

Dark CPT steady states decouple from the two fields in either $\Lambda$ or V configurations and cause a corresponding EIT window in the ensemble absorption. Absorption and dispersion manipulation by such electromagnetically induced transparency (EIT) is the operational principle of EIT quantum memories. In \cref{sec:memory} we consider the prospects for $T$ centre EIT memories (both optical and microwave), but first we perform one final preparatory measurement. In the same $\Lambda$ or V configuration, but in the limit of increasing pump power, the CPT window transitions to an Autler-Townes split transition, which is itself the basis of an Autler-Townes quantum memory.

\section{Autler-Townes splitting}
\label{sec:ATS}

The operational bandwidth of EIT quantum memories is limited by the two-photon transition linewidth. Autler-Townes (AT) quantum memories utilize selective absorption by the spectrally distinct AT-split dressed states of a single transition to circumvent this limit \cite{Saglamyurek2018a}. With a modified ODMR scheme we can observe AT splitting of MW transitions in the same $T$ ensemble. For this measurement a slightly smaller magnetic field is chosen, $B_0 = 60$~mT along the [110] axis of the sample. 

\begin{figure}[]
\includegraphics[width=\linewidth]{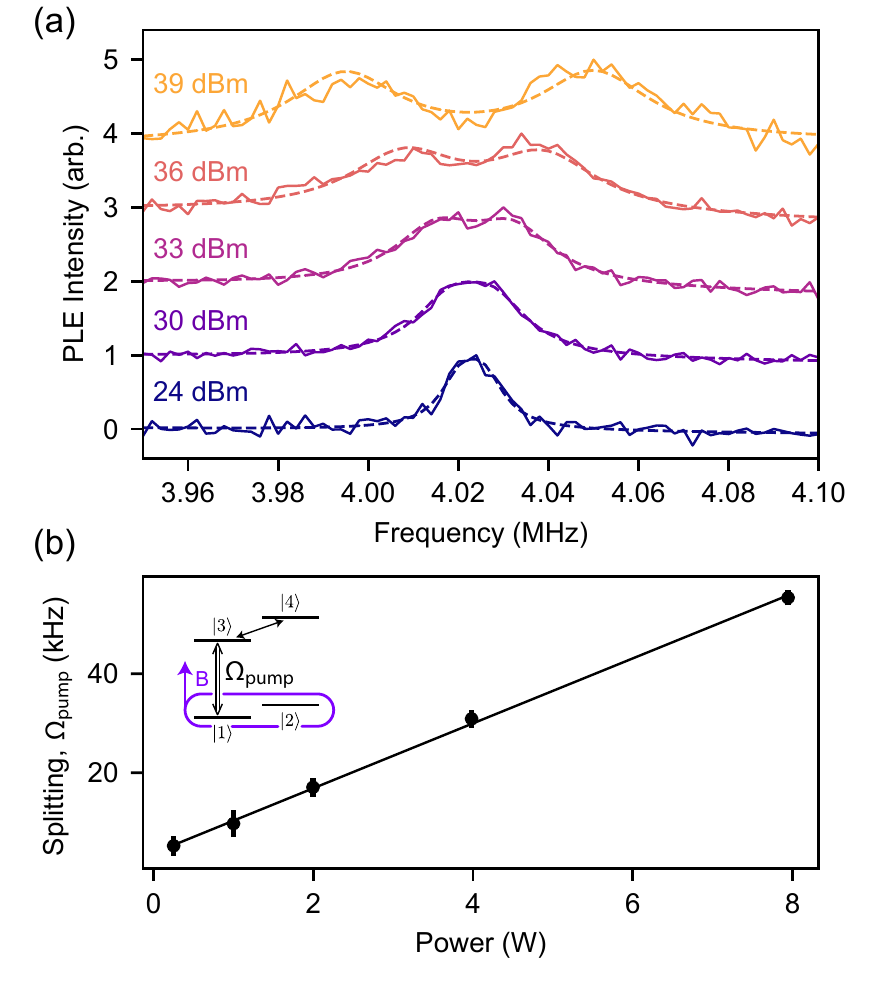}
\caption{Autler-Townes splitting of the MW$_\Downarrow$ transition as a function of MW$_\Downarrow$ power. (a) Spectra taken by scanning a non-phase coherent RF source over the nuclear flip transition. ODMR PL signal is generated by exciting to the TX$_0$ state by a laser resonant with the B$_1$ optical transition (scheme shown inset in (b)). As the MW$_\Uparrow$ power (text labels) increases, AT splitting becomes clear. Dashed lines are a split-peak fit to each dataset. (b) Splitting from fits (equal to the MW$_\Uparrow$ Rabi frequency $\Omega_\mathrm{pump}$) as a function of resonant power.}
\label{fig:4}
\end{figure}

AT dressed states are split by the pump Rabi frequency $\Omega_\mathrm{pump}$. Rabi frequencies fitted to the CPT notches above, and observed by EPR with the same apparatus \cite{Bergeron2020}, indicate that attainable AT splittings $\Omega_\mathrm{pump} < 0.2$~MHz will not be clearly resolved with the ODMR schemes above. Instead, we apply an alternative ODMR scheme in which a radio frequency (RF) probe field scans over one of the NMR transitions. By this method we isolate a single sub-ensemble and achieve ODMR linewidths narrow enough to measure the transition to AT splitting as $\Omega_\mathrm{pump}$ increases. \Cref{fig:4}(a) shows ODMR spectra of the transition $\ket{3}$--$\ket{4}$ as the MW power resonant with MW$_\Uparrow$ increases. The RF probe field is swept across the $\ket{\uparrow_\textrm{E}}$ nuclear transition between states $\ket{3}$ and $\ket{4}$. As the MW pump power increases, the RF resonances split by $\Omega_\mathrm{MW}$, the microwave Rabi frequency. 
We measure AT splitting up to $55(1)$~kHz, demonstrating the basic mechanism required for microwave AT quantum memories. Splitting is limited in this case by the available MW power and inefficient resonator design. Larger splittings for improved optical and MW memory bandwidths are possible. In the following section we compare EIT and AT memory models to assess the prospects of $T$ centre memories for microwave and optical fields, and determine the optimal bandwidth for operating $T$ centre memories at feasible optical depths.

\section{Free-space optical memory }
\label{sec:memory}

The coherence times of the $T$ centre electron and nuclear spins, $<2.1$~ms and $<1.1$~s respectively, are appealing for quantum memory in either optical quantum networks or quantum processors. It remains to determine which of the many quantum memory schemes are best suited to the optical and microwave properties of $T$ centre ensembles determined in the literature and in the measurements above. Storage of a weak signal (i.e., probe) pulse using a relatively strong control field in a $\Lambda$-type system has been studied using different protocols such as EIT, off-resonant Raman, and ATS \cite{lukin2003colloquium,Lvovsky2009,reim2010towards,Saglamyurek2018a}.

Keeping in mind that the nuclear splitting is not optically resolvable, there are four optical transitions per $T$ centre orientation that can be addressed individually and formed into two possible optical $\Lambda$ configurations addressing the electron spin coherence. In the ideal lifetime-limited case a $T$ centre optical quantum memory could directly address the longer-lived nuclear spin coherence, as in the MW memory proposal below, for storage times beyond $1$~s. However even the electron spin coherence time of $\approx2$~ms is sufficient for a basic repeater demonstrations and swapping between the electron and nuclear spins is possible. 

In this section we consider theoretically the $\Lambda$-type system comprising two long-lived ground states $\ket{g_1} = \ket{\uparrow_e}$ and $\ket{g_2} = \ket{\downarrow_e}$ that are optically connected to a common excited state $\ket{\downarrow_h}$.  We assume both electron and nuclear spins are initially polarized. We then use a strong control field with the Rabi frequency of $\Omega$ in resonance with the optical transition $B$, to store a weak signal pulse coupled to the $A$ transition (see \cref{fig:1}). 

Based on the optical depth measurement performed in \cref{sec:OD}, $T$ centre ensembles could be made with optical depth sufficient for resonant optical quantum memory schemes given feasible concentration improvements. However, absent cavity-enhanced absorption, even these optical depths are insufficient for off-resonant schemes including gradient echo memory (GEM). Resonant memory schemes that efficiently utilize the available optical depth are best suited to near-term $T$ centre ensembles. In the following subsections we consider in detail the prospects of EIT and ATS free-space optical quantum memory schemes. Although \TwoEightSi{} $T$ centre ensembles exhibit optical inhomogeneous broadening of at least 33 MHz \cite{Bergeron2020}, throughout this
analysis we consider spectral tailoring of the ensemble bounded from below by the homogeneous linewidth $\Gamma/2\pi$. Therefore, we neglect the effect of inhomogeneous broadening in our estimations.

\subsection{EIT Memory}

In the EIT protocol, ramping down the control field power gradually to zero allows the adiabatic transfer of signal coherence to the spin-wave mode. EIT can be implemented effectively in both narrowband, i.e. $B_{\text{sig}}<\Gamma/2\pi$, and broadband, i.e. $B_{\text{sig}}>\Gamma/2\pi$, signal regimes where $B_{\text{sig}}$ is the bandwidth of the signal at full-width half-maximum, and $\Gamma/2\pi$ is the linewidth of the signal transition. However, this protocol is best suited for the narrowband regime where control field optimization is achievable at any optical depth. Eliminating signal absorption and maintaining adiabatic evolution with a broad transparency window requires both a very large optical depth and a strong control field \cite{wei2020broadband, rastogi2019discerning}.

In general, EIT efficiency optimization is achievable through control field and/or signal pulse optimization \cite{gorshkov2007freespace,novikova2007optimal,novikova2008optimal}. The consistency of these optimization methods has been demonstrated in Ref. \cite{phillips2008optimal}. In the narrowband regime and for $d>20$, control field optimization is achievable by keeping the process adiabatic while satisfying the condition $\tau_d/\tau_{\text{sig}}\approx2$, where $\tau_d=d\,\Gamma/\Omega^2$ is the group delay and $\tau_{\text{sig}}$ is the signal time at FWHM \cite{rastogi2019discerning}. Considering the probe field with a Gaussian temporal profile, the control optical Rabi frequency $\Omega$ should therefore satisfy the condition $\Omega^2 \approx d \,\Gamma B_{\text{sig}} /0.88$.

For the $T$ centre with $d=27$, corresponding to feasible near-term optical depths as discussed in \cref{sec:OD}, the maximum achievable efficiency of the EIT protocol for the forward retrieval can be estimated as $\sim71\%$ \cite{phillips2008optimal}. The memory efficiency for the backward propagation depends on the ground states splitting $ \omega_{g_1 g_2}$ (as the non-zero splitting breaks the conservation of momentum in backward
retrieval \cite{gorshkov2007freespace}). Therefore, for a non-zero splitting, the backward efficiency could be lower than the efficiency of the forward retrieval unless we make $\sqrt{d}\gg L \omega_{g_1 g_2} /c$ where $L$ is the length of the medium and $c$ is the speed of light.

This EIT memory efficiency estimate applies for short storage times. Reduction of the efficiency due to the spin decoherence can be taken into account by adding a term $\text{exp}(-\gamma_s t)$ to the efficiency during the storage time \cite{phillips2008optimal,gorshkov2007freespace}.
Other imperfections such as four-wave mixing \cite{phillips2008optimal, lauk2013fidelity} may further reduce the experimentally achievable efficiency.

\subsection{ATS Memory}
\label{ssec:ATS-memory}

Unlike the EIT protocol, which relies on the adiabatic elimination of the atomic polarization mode, in ATS memory polarization mediates the non-adiabatic coherence exchange between the signal pulse and spin modes. In the broadband regime, ATS is less demanding in terms of technical requirements such as optical depth and control field power compared to the EIT memory. However, in the narrowband regime, the efficiency of the ATS memory is close to zero as the average coherence time of the transitions in resonance with the signal and control fields becomes shorter than the interaction time (i.e., signal pulse duration).

Assuming the system is initially prepared in the state $\ket{g_1}$, we can use dynamically controlled ATS lines produced by a strong
control field that drives the $\ket{g_2}$--$\ket{e}$ transition to absorb
a weak signal pulse in resonance with the $\ket{g_1}$--$\ket{e}$ transition \cite{Saglamyurek2018a}. 
Absorption of the signal pulse by ATS peaks (with a peak separation equal to the control Rabi frequency) will then map its coherence to a collective state between $\ket{g_1}$ and $\ket{e}$ (i.e., polarization mode), which subsequently evolves into a collective spin-excitation between the ground states. After the write (i.e., storage) process, we abruptly switch the control field off to trap the coherence, wait for storage time, and turn it on again for the read-out (i.e., retrieval) process. Using the control-field optimization, one can increase the spectral overlap between the ATS peaks separation and the signal bandwidth to maximize the signal pulse absorption. Here, the control-field optimization requires the control pulse area for both write and read stages to be $2\pi$ \cite{rastogi2019discerning}. Using the ATS memory protocol for the $T$ centre, a relatively high efficiency is achievable for broadband light storage and retrieval.

\begin{figure}[]
\includegraphics[width=\linewidth]{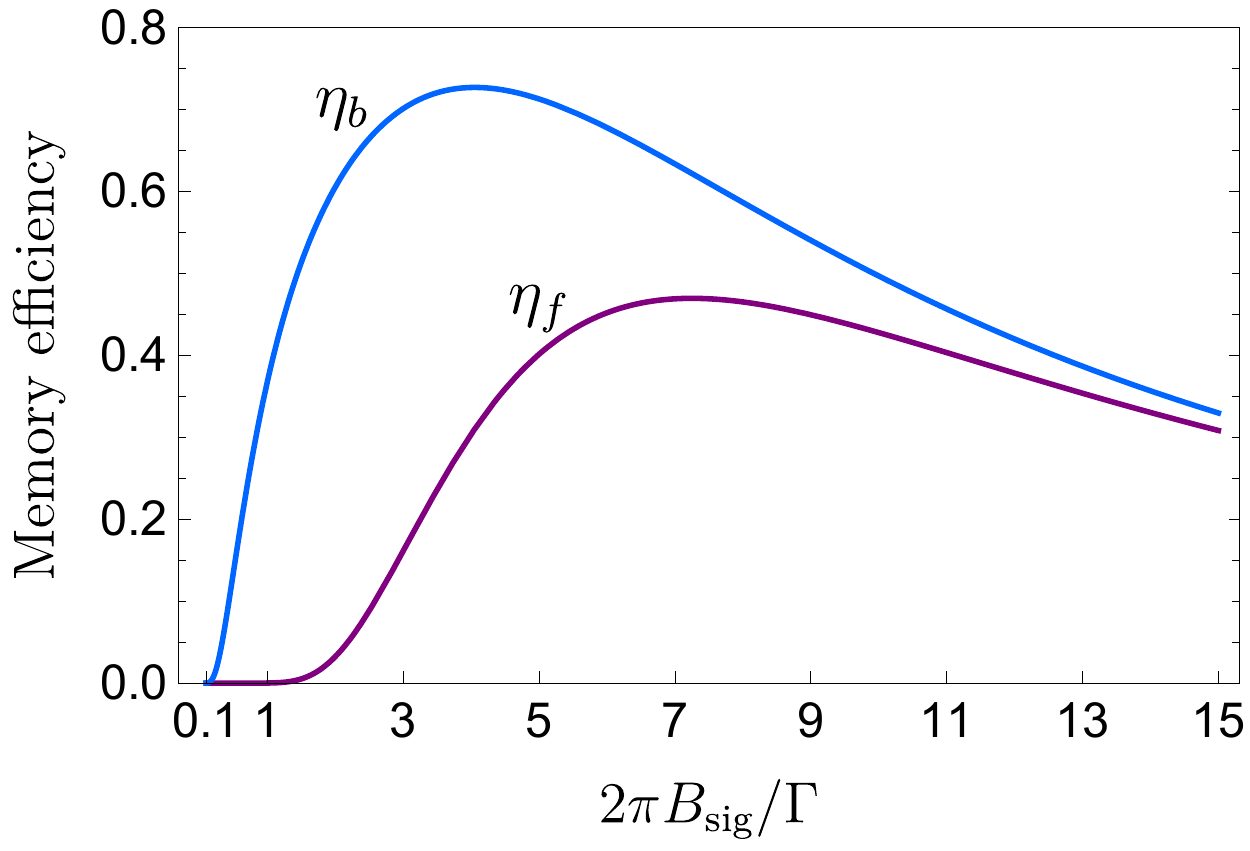}
\caption{Free-space forward $\eta_f$ and backward $\eta_m$ memory efficiency with respect to the ATS factor for $d=27$ in the broadband regime.}
\label{fig:ATS}
\end{figure}

In Ref \cite{saglamyurek2018coherent}, ATS memory efficiency in the free space has been discussed in terms of the ATS factor $F=\Omega/\Gamma$ (or equivalently $F=2\pi B_{\text{sig}}/\Gamma$).  
As shown in Fig.\ref{fig:ATS}, for the $T$ centre with $d=27$, 
 the maximum efficiency of $\eta_{\text{b},\text{max}}=72.6\%$ ($\eta_{\text{f},\text{max}}=46.9\%$) for $F=4$ ($F=7.25$) is achievable in the backward (forward) propagation. For the optical memory,  assuming $\Gamma/2\pi=27$ MHz \cite{MacQuarrie2021}, these correspond to the Rabi frequency of $\Omega=2\pi\times108$ and $2\pi\times196$ MHz for the backward and forward modes, respectively.

\section{Microwave memory}\label{sec:mw}

We now turn our attention to $T$ centre prospects for MW quantum memories. In \cref{sec:CPT,sec:ATS} we have measured coherent MW interactions in $T$ centre ensembles from the CPT to AT regimes. MW AT splitting such as we observed is a precursor to the demonstration of a $T$-ATS MW memory. We will treat such an ATS memory theoretically to determine the prospects of MW memories with near term $T$ centre ensembles. For this study of the MW ATS memory we choose one of the two possible MW $\Lambda$ schemes shown in \cref{sec:CPT}, assigning $\ket{1}$ and $\ket{2}$ as ground states $\ket{g_1}$ and $\ket{g_2}$, respectively, and $\ket{3}$ as excited state $\ket{e}$. We begin with an initial state $\ket{g_1}$, and use dynamically controlled ATS lines produced by a control field,  as observed in \cref{sec:ATS}, to store a signal field resonant with the MW$_\Uparrow$ transition (see Fig.\ref{fig:1}). So that the coherence of the signal pulse is mapped into the collective nuclear spin excitations of the atoms. The theoretical treatment from Ref. \cite{saglamyurek2018coherent} applies to the ATS MW memory as well as the optical memory considered above. For this case lower MW transition linewidths, as measured in \cref{sec:CPT}, yield lower $\Omega$ for the MW ATS memory.

However in general the weakness of magnetic dipole transitions requires a MW resonator to enhance the MW-ensemble coupling strength. To describe this resonantly enhanced interaction we can use the cavity input-output formalism i.e, $\hat{E}_\text{out}=-\hat{E}_\text{in}+\sqrt{2\kappa}\hat{E}$, and Heisenberg equations of motion \cite{Saglamyurek2018a,gorshkov2007photon}
\begin{equation}
\begin{aligned}
\dot{\hat{P}}&=-\gamma_e \hat{P}+i g(t)\sqrt{N}\hat{E}+\frac{i}{2} \Omega \hat{S},\\
\dot{\hat{S}}&= \frac{i}{2} \Omega^* \hat{P},\\
\dot{\hat{E}}&= i g(t)\sqrt{N}\hat{P}-\kappa\hat{E}+\sqrt{2\kappa}\hat{E}_\text{in},
\end{aligned}
\label{Hem}
\end{equation}
where $\hat{P}$ and $\hat{S}$ are the polarization and spin-wave operators, $\hat{E}$ is the cavity field which we assume can be adiabatically eliminated, 
$\kappa$ is the cavity decay rate, $\gamma_e=\Gamma/2$ is the decoherence rate of the polarization mode 
and $g(t)$ is the time-dependent light-matter coupling per emitter and $N$ is the number of atoms in the ensemble. Note that here we have ignored the Langevin noise operators (i.e., the incoming noise is vacuum) and once again assumed the spin-wave decoherence rate is negligible as per \cref{sec:memory}. 

\begin{figure}[]
	\includegraphics[width=\linewidth]{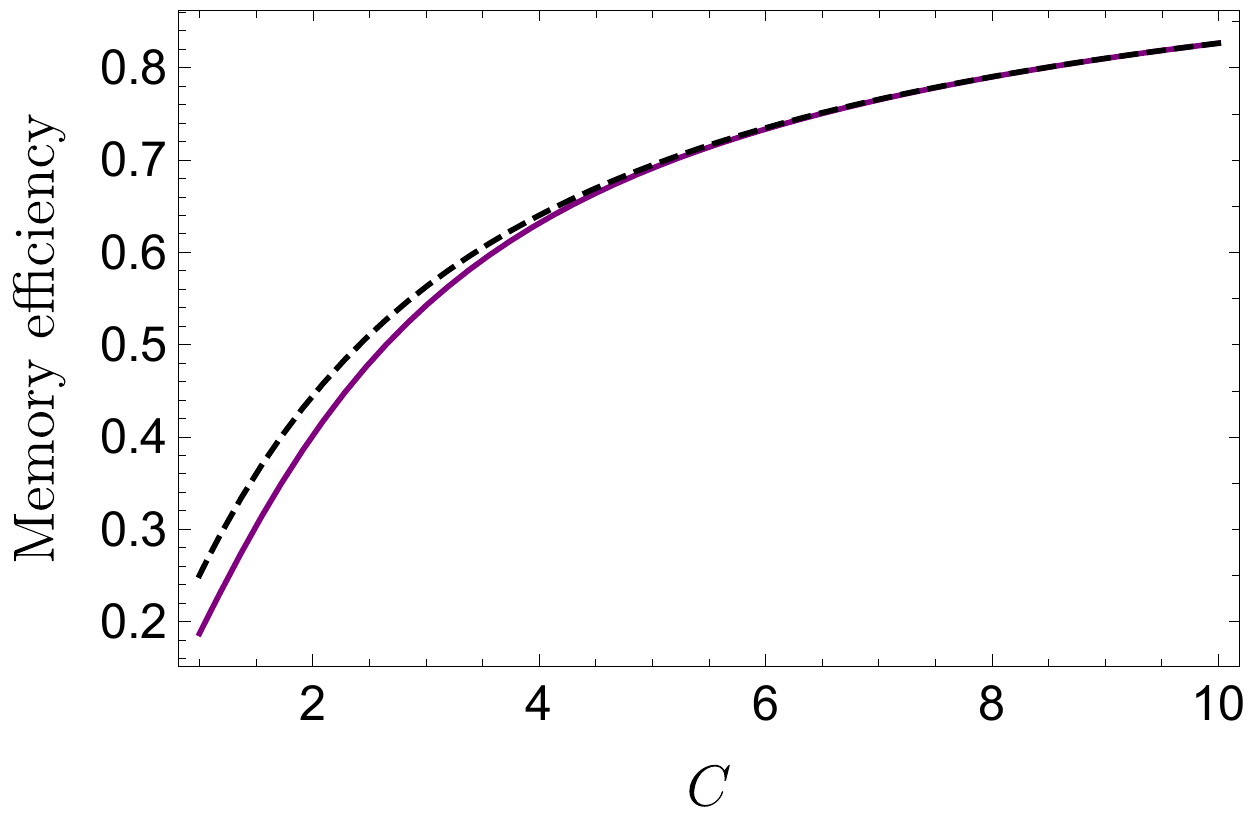}
	\caption{Memory efficiency as a function of the resonator cooperativity. The solid line shows the efficiency given by equation (\ref{eq:eff}) assuming $t_s=t_r=1/(2\gamma_{e})$. The dashed line corresponds to the $t_r\rightarrow \infty$ (i.e., $\eta=C^2/(1+C)^2$).}
	\label{fig:transductionefficency}
\end{figure}

To estimate the overall efficiency of the memory in the presence of a resonator, we follow the same optimal scheme as in Ref \cite{gorshkov2007photon} for the non-adiabatic or fast limit $\Omega> \Gamma C$, where $C$ is the cooperativity parameter $C=Ng^2/\kappa\gamma_{e}$. The overall efficiency is $\eta=\eta_s\eta_r $, the product of the optimal storage efficiency $\eta_s$ (average ratio of stored excitations to incoming photons) and the optimal retrieval efficiency $\eta_r$ (average ratio of re-emitted photons to stored excitations). 
In the optimal regime, we compute the retrieval efficiency as 
\begin{equation}
\eta_r=\int_{0}^{t_r}dt\frac{2N}{\kappa}g^2(t)e^{-\int_{0}^{t}dt^\prime 2(Ng^2(t^\prime)/\kappa\,+\,\gamma_{e})},\label{eta_r}
\end{equation}
where $t_r$ is the effective time elapsed during the retrieval process. For simplicity, it is assumed that the retrieval process starts at $t=0$ rather than at some time $t_s>\tau$ where $\tau$ is the duration of the signal, and that $S(0)=1$ and $P(0)=0$.
It has been shown that using the optimal strategy the storage and retrieval efficiencies are analogous \cite{gorshkov2007universal, heshami2012controllable, gorshkov2007photon}. Hence, in the optimal regime, the overall efficiency can be estimated
\begin{equation}
\eta \simeq\frac{C^2}{(1+C)^2} \left(1-e^{-2\gamma_{e}(1+C)t_s}\right)\left(1-e^{-2\gamma_{e}(1+C)t_r}\right).
\label{eq:eff}
\end{equation}
In deriving Eq.\ref{eq:eff}, it is assumed that during the retrieval process all spin mode excitations have been evolved back to the polarization mode (i.e., no excitations are left in the spin mode). For fixed and finite values of the effective times, as the cavity cooperativity increases, the exponentially decaying terms of the Eq.\ref{eq:eff} go to zero, and hence, the first term dominates. Equivalently, in Eq.\ref{eta_r}, if we initially assume $t_r\rightarrow \infty$ such that no excitations are left in the T ensemble i.e., $P(\infty)=0$, the optimal retrieval efficiency  reduces to $C/(1+C)$ and the overall efficiency becomes  $C^2/(1+C)^2$ (see Fig.\ref{fig:transductionefficency}) \cite{gorshkov2007photon}. With this maximum possible efficiency, one can store photons with $\tau\approx 1/(C\gamma_{e})$ where the polarization decay rate is negligible.

From this analysis we can see that resonator cooperativities of order $10$ are required to exceed $80$\% memory efficiency with a $T$ centre ensemble. Although we have introduced them in the context of the MW memory scheme, Eqs.\ref{Hem} apply to resonantly enhanced optical memories as well. In the following section we will propose to combine resonantly enhanced MW and optical memories to realize a $T$ centre MW to optical transducer with $T$ centre ensembles.

\section{Microwave-to-optical transduction}
\label{sec:transduction}

Motivated by the optical and microwave memory performance estimates above, we will now look at combining the capabilities to form a $T$ centre ensemble microwave-to-optical transducer that stores MW photons as spin-wave excitations and recalls them optically. For this operation we will consider both resonantly enhanced optical and microwave ATS memories.

Transduction may be achieved by way of two $\Lambda$ schemes, optical and microwave, addressing the same spin coherence. If this can not be achieved, it is necessary to transfer the coherence by an intermediate step. In the following, we elaborate on three possible approaches.

First, we may consider the MW and optical $\Lambda$ schemes set out in \cref{sec:memory} and \cref{sec:mw}. The MW configuration utilizes the long-lived nuclear spin coherence $\rho_{12}$, however, as discussed in \cref{sec:memory}, the hyperfine splitting between $\ket{1}$ and $\ket{2}$ is not optically resolved in $T$ centre ensembles measured to date. Transduction from this microwave interface to an optical field using the same optical $\Lambda$ scheme we utilized in \cref{sec:memory} for an optical AT memory requires one to (i) store MW photons in resonance with the $\ket{1}-\ket{3}$ transition as nuclear spin-wave excitations between the states $\ket{1}$ and $\ket{2}$, (ii) transfer the nuclear spin wave $\rho_{12}$ to an electron spin wave $\rho_{13}$ and (iii) recall the stored pulse using a control field in resonance with the optical transition $B_1$. Transferring coherence between the nuclear and electron spin waves (step (ii)) requires only coherent control of the $T$ centre ground states, which has already been demonstrated over ensembles in Ref.~\cite{Bergeron2020}.

Alternatively, one may consider schemes that avoid step (ii) above. One option is to use the electron spin for MW storage via a different $\Lambda$ configuration i.e. $\ket{1}$ and $\ket{3}$ as the ground states, and $\ket{4}$ as the excited state. This configuration allows us to store MW photons in resonance with the $\ket{1}$--$\ket{4}$ transition while the control field is resonant with the $\ket{3}$--$\ket{4}$ transition. Note that the decoherence rate of the $\ket{1}$--$\ket{3}$ and $\ket{1}$--$\ket{4}$ transitions are comparable. In general this results in ATS line broadening and reduced efficiency \cite{Saglamyurek2018a}. However, in the limit that electron spin coherence time is very long, as it is in the $T$ centre \cite{Bergeron2020} such ATS line broadening is negligible. 

Finally, the transfer step is unnecessary when the hyperfine splitting of the nuclear spin wave is optically resolved. For example, in a scheme that utilizes larger hyperfine splittings due to nuclear spins at the carbon sites in $^{13}$C isotopic variants of $T$ \cite{Bergeron2020}, at higher magnetic fields, or by improving the optical linewidth such that the hydrogen hyperfine splitting is resolved. This way one can consider $\ket{1}$ and $\ket{2}$ as the joint ground states, and $\ket{3}$ and $\ket{\downarrow_\mathrm{h}}$ as the excited state of the MW and optical $\Lambda$ systems, respectively. 

\subsection{Transduction efficiency and fidelity}
Optimal efficiency of the transducer, where MW and optical transitions are coupled to the respective cavities, can be estimated using the same optimal strategy as for the ATS memory except that there is now an additional mode mismatch factor $\eta_m$ that should be taken into account. As a result, the overall efficiency would be $\eta=\eta_s\eta_r\eta_m$. The mode mismatch factor depends on the design (geometry) of the cavities. In the ideal case, where all atoms are placed in the maximum of both optical and MW fields, the mode mismatch factor can reach unity. In this case and for $t_r\rightarrow \infty$, the optimal efficiency of the transducer is estimated as $C_sC_r/(1+C_s)(1+C_r)$, where $C_s$ and $C_r$ are the cooperativity of the MW and optical cavities, repetitively. 

The other metric to quantify the quality of transduction is the overlap between the transduced signal and the transduced noises. We refer to this measure as the transduction fidelity. In order to compute the fidelity, one could solve the system dynamics by taking several important imperfections such as spin decoherence, microwave thermal photons, and inhomogeneous broadening into account to determine the fidelity. As discussed before, inhomogeneous broadening can be dealt with by spectral tailoring. Therefore, here we only consider the thermal photons in the microwave cavity. Spin decoherence is also negligible given that the process can be much faster than the spin coherence times \cite{Bergeron2020}. For this calculation, we consider a transduction system consisting of $\ket{1}$ and $\ket{2}$ as the ground states, and $\ket{3}$ and $\ket{\downarrow_h}$ as the excited state of the microwave and optical $\Lambda$ subsystems, respectively.

The initial microwave thermal occupation $\bar{n}_{\text{th}}=1/(e^{(\hbar\omega_{1, 3}/k_B T)}-1)$ is estimated to be $0.0045$ with $T=20$ mK and $\omega_{1, 3}=2\pi\times 2.25$ GHz \cite{Bergeron2020} which is the splitting between $\ket{1}$ and $\ket{3}$. Since the initial microwave photon occupation is quite small, and it does not affect the quantum system dynamics significantly, we can treat it separately from the signal. The fidelity can be computed using the signal-to-noise ratio (SNR) \cite{asadi2022proposal}:
\begin{equation}
F=\frac{1}{1+\text{SNR}^{-1}}.
\label{eq:fid}
\end{equation}
Given that the thermal noise and signal can be treated independently, it is valid to compute the SNR even before the transduction process starts. In this way, the SNR is given by:
\begin{equation}
\text{SNR}=\frac{n_{\text{sig}}}{n_{\text{th}}},    
\end{equation}
where $n_{\text{sig}}$ is the mean occupation of signal in microwave cavity. For a single-photon signal, $n_{\text{sig}}=B_{\text{sig}}/\kappa_s$, where $\kappa_s$ is the microwave cavity decay rate. This is valid in the regime that $B_{\text{sig}}\ll\kappa_s$. Given that $n_{\text{th}}\ll n_{\text{sig}}$, the fidelity shown in Eq.(\ref{eq:fid}) can be simplified as:    
\begin{equation}
F\approx 1-\frac{n_{\text{th}}}{n_{\text{sig}}}=1-\frac{n_{\text{th}}\kappa_s}{B_{\text{sig}}}.    
\end{equation}
Since we are interested in the case where the transduction efficiency is maximized, $B_{\text{sig}}$ can be expressed as $B_{\text{sig}}=C_s\gamma_{e_s}$. Hence, in the regime of maximum efficiency, the fidelity is explicitly given by:
\begin{equation}
F\approx 1-\frac{n_{\text{th}}\kappa_s}{C_s\gamma_{e_s}}.  
\label{eq:ff}
\end{equation}
Clearly, as the cavity cooperativity increases, the fidelity increases, and the efficiency also increases as shown in Fig. \ref{fig:transductionefficency}.

The other way to compute the fidelity is to estimate the SNR after the transduction, which should be equivalent to the approach discussed above. In this way, we can model the microwave thermal photons as dark counts \cite{Ji2022proposalroom}. The dark count rate is given by $D=R\eta$ with the transduction efficiency $\eta$ and the average rate of thermal photons $R$. It describes the effective number of photons emitted as background noise in the optical cavity per unit of time. The probability distribution associated with this dark count rate is given by a Poisson distribution $P(T_{\text{tr}},D)=D^nT^n_{\text{tr}}e^{-DT_{\text{tr}}}/n!$ where $n$ is the number of dark counts and $T_{\text{tr}}$ is the total transduction time. Thus, this time can also be written as $T_{\text{tr}}=t_s+t_r$.

Then, for a single-photon transduction, the SNR takes the following form:
\begin{equation}
\text{SNR}=\frac{\eta n_{\text{sig}}}{\overline{N}_D},    
\end{equation}
where $\overline{N}_D$ is the mean number of dark counts. Specifically, at a given time $T_{\text{tr}}$ the SNR becomes:
\begin{equation}
\text{SNR}=\frac{\eta n_{\text{sig}}}{DT_{\text{tr}}},    
\end{equation}
where $DT_{\text{tr}}$ is the mean number of dark counts that follow the Poisson distribution. It can be further simplified as $\text{SNR}=n_{\text{sig}}/RT_{\text{tr}}$. In the regime of the maximum efficiency, $RT_{\text{tr}}=n_{\text{th}}$, we recover $\text{SNR}=n_{\text{sig}}/n_{\text{th}}$, which leads to the fidelity shown in Eq.(\ref{eq:ff}).  

\section{Discussion} \label{sec:discussion}

The $T$ centre appears to combine several properties appealing for both microwave and quantum memories, and transduction between them, including long lived spins and telecom optical emission. We have considered possible quantum memory and transduction schemes for ensembles of $T$ centres, and  performed preliminary measurements with ensembles in \TwoEightSi{} to establish properties informing memory and transduction device design. 

Absorption measurements indicate a resonant optical depth of $0.009$ in a narrow-linewidth ensemble. Cautiously forecasting a small improvement over $T$ centre concentrations that have already been demonstrated in devices, we project that an optical depth of $27$ with a similar sample is possible. This optical depth is sufficient for efficient optical quantum memory by EIT or ATS with estimated a memory efficiency as high as $73\%$.

In addition to the potential for optical memory, the $T$ centre's optical ground state spin structure features $\Lambda$ and V configurations suitable for storing MW fields. We demonstrated these configurations by ODMR spectroscopy, and measured CPT windows in $\Lambda$ and V configurations as a precursor to an EIT memory for MW fields. Fluorescence SNR requirements prohibit us from reaching the low-power limit, but even so we measured CPT linewidths of $17(2)$~kHz, significantly below the ODMR linewidth, and resolved orientation sub-ensembles that were degenerate to MW ODMR precision. We demonstrated microwave ATS up to $55(1)$~kHz splitting. We then discussed a cavity-enhanced MW memory protocol and arrived at an efficiency estimate as a function of cavity cooperativity. In particular, MW memory efficiencies of greater than $80\%$ are possible for $C=10$.

In general, the operation of MW and optical $T$ centre quantum memories can be combined to transduce between the microwave and optical fields. Transduction requires either (i) an optical quantum memory addressing the nuclear spin coherence $\rho_{12}$, (ii) a microwave memory addressing an electron spin coherence such as $\rho_{13}$ via an NMR control field, or (iii) coherent transfer between $\rho_{12}$ and $\rho_{13}$ at an intermediate stage. (iii) requires only the coherent control of $T$ ensembles that has already been experimentally demonstrated \cite{Bergeron2020}. Furthermore, (i) is within reach of schemes utilizing the known $^{13}$C $T$ centre variants.

We proposed a transduction protocol where both microwave and optical transitions are coupled to resonators, and discussed the efficiency and fidelity of the system. Mutually compatible MW and optical cavities with $C = 10$ would yield microwave to optical transduction efficiencies of $83\%$. Necessary future work includes further improving ensemble concentrations, and engineering optical and microwave resonators meeting these metrics, but an in-principle $T$ centre transduction demonstration could be achieved with designs available in the literature. 

We have demonstrated the potential of this new spin-photon platform for quantum memory, and identified pathways towards efficient microwave-to-optical transduction with $T$ centres. Further developing and transplanting this approach into on-chip devices, for example with integrated silicon photonic resonators combining moderate mode confinement and high quality factors, would yield an on-chip spin-photon interface for quantum memories and microwave to optical transducers. These devices could network directly between quantum information platforms that share the same technologically and commercially advanced silicon chip platform (including superconducting qubits, quantum dots, trapped ions, and silicon impurity spins) and the telecom optical fibres of a global quantum internet.

\section*{Acknowledgements}
This work was supported by the Natural Sciences and Engineering Research Council of Canada (NSERC), the New Frontiers in Research Fund (NFRF), the National Research Council (NRC) of Canada through its High-Throughput Secure Networks (HTSN) challenge program, the Canada Research Chairs program (CRC), the Canada Foundation for Innovation (CFI), the B.C. Knowledge Development Fund (BCKDF), and the Canadian Institute for Advanced Research (CIFAR) Quantum Information Science program. D.B.H is supported by the Banting Fellowship program.

The $^{28}$Si samples used in this study were prepared from the Avo28 crystal produced by the International Avogadro Coordination (IAC) Project (2004--2011) in cooperation among the BIPM, the INRIM (Italy), the IRMM (EU), the NMIA (Australia), the NMIJ (Japan), the NPL (UK), and the PTB (Germany). We thank Alex English of Iotron Industries for assistance with electron irradiation.

\section*{APPENDIX A: SAMPLE PREPARATION} \label{sec:Sample}

All measurements are performed on one of two \TwoEightSi{} crystals, previously described in Ref. \cite{Bergeron2020}. These have been purified to 99.995\% \TwoEightSi{} isotope proportion by the Avogadro project and then electron irradiated and thermally annealed to produce $T$. In addition to isotopic purity these samples have low chemical impurity, with less than $10^{14}$~oxygen/cm$^3$. $T$ centres are formed from naturally present low-level carbon impurities, which differ between the two samples. The sample used for absorption measurements has $1.5\times10^{15}$~carbon/cm$^3$, the sample used for CPT and ATS measurements has  $5\times10^{14}$~carbon/cm$^3$. 

\section*{APPENDIX B: METHODS}\label{sec:Methods}

\textit{Cryogenics} 
\vspace{2mm}\newline
The samples are loosely mounted in a liquid Helium immersion dewar such that the crystal is not strained. The temperature of the Helium-4 bath is adjusted between $1.4$--$4.2$~K by pressure control.

\vspace{5mm}
\textit{Photoluminescence spectra}
\vspace{2mm}\newline
The sample is  illuminated with an above-bandgap $1047$~nm laser and the resulting broadband photoluminescence is directed into a Bruker IFS 125 HR Fourier Transform infrared (FTIR) spectrometer with a CaF$_2$ beam splitter and liquid nitrogen cooled Ge photodetector.

\vspace{5mm}
\textit{Absorption measurement}
\vspace{2mm}\newline
The sample is illuminated along its longest side by a broadband light source with beam size smaller than the sample cross section. Light passes through the sample and two cryostat windows before being directed into the Bruker FTIR spectrometer. The light source spectrum is free of structure in the vicinity of the $T$ centre absorption lines and is removed by subtracting a linear fit about the absorption dip location.

\vspace{5mm}
\textit{Photoluminescence excitation}
\vspace{2mm}\newline
Photoluminescence excitation spectra are recorded by scanning a narrow, tunable laser over the TX$_0$ ZPL and detecting lower energy photons emitted into the TX phonon sideband. Resonant excitation is performed with a Toptica DL100 tunable diode laser. The optical power is amplified up to $100$~mW with a Thorlabs BOA1017P and the beam is expanded to diameter $2$--$4$~mm. Non-resonant excitation light due to spontaneous emission in the laser and amplifier is removed by a combination of an Edmund Optics 87-830 $1350 \pm 12.5$~nm and Iridian Spectral Technologies DWDM $1329\pm0.5$~nm bandpass filters. The two filters are tilted so as to tune them onto the TX$_0$ resonance.

Stray excitation light is removed from the detection path by two Semrock BLP02-1319R-25 $1319$~nm longpass rejection filters, which in fact block the excitation light at TX$_0$ with OD$=4.5$ at normal incidence, but pass light with wavelength longer than $1350$ nm. A $1375$($50$)~nm bandpass filter further selects a detection window in the TX phonon sideband that excludes a silicon Raman replica of the pump laser at $1426$~nm. The fluorescence rate in this detection window is measured by an IDQuantique ID230 avalanche photodiode with 25\% quantum efficiency. 

\vspace{5mm}
\textit{Optically Detected Magnetic Resonance}
\vspace{2mm}\newline
Magnetic resonance experiments are performed with the sample in an adjustable magnetic field applied by an iron core electromagnet within the PLE apparatus as described above. The sample is held at the centre of a split-ring resonator with a resonant frequency of $2.25$~GHz and $10$~MHz bandwidth. MW signals from by two SRS SG384 and SG386 signal generators are combined  with a ZB2PD-63-S+ power splitter and then amplified up to 1W of power by ZHL-16W-43-S+ and ZHL-1-2W+ amplifiers as required. 
\FloatBarrier
%

\end{document}